\documentclass{aa}  
\usepackage{graphicx}
\usepackage{longtable}
\usepackage{multirow}
\usepackage{rotating}
\usepackage{natbib}
\usepackage{amsmath}
\usepackage{txfonts}

\newcommand{\um}{$\mu$m}                                 
\newcommand{\lsim}{\;\lower.6ex\hbox{$\sim$}\kern-7.75pt\raise.65ex\hbox{$<$}\;}
\newcommand{\gsim}{\;\lower.6ex\hbox{$\sim$}\kern-7.75pt\raise.65ex\hbox{$>$}\;}
\newcommand{\gl}{\;\lower.6ex\hbox{$<$}\kern-7.75pt\raise.65ex\hbox {$>$}\;}
\newcommand{\asec}{$^{\prime \prime}$}
\newcommand{\adeg}{$^{\circ}$}

\newcommand{\dder}{$\partial ^2$}
\newcommand{\sigdder}{$\sigma_{\partial ^2}$}
\begin{document}
\title{Source extraction and photometry for the far-infrared and sub-millimeter continuum in the presence of complex backgrounds}


\author{S. Molinari
	\inst{1} \and
	E. Schisano
	\inst{1} \and
	F. Faustini
	\inst{2} \and
	M. Pestalozzi
	\inst{1} \and
	A.M. Di Giorgio
	\inst{1} \and
	S. Liu
	\inst{1}
          }


   \institute{INAF-Istituto Fisica Spazio Interplanetario, Via Fosso del Cavaliere 100, I-00133 Roma, Italy 
              \email{molinari, schisano, pestalozzi, digiorgio@ifsi-roma.inaf.it} \and
             ASI Science Data Center, Via E. Fermi, Frascati, Italy 
             \email{faustini@asdc.asi.it} 
             }

   \date{Received ; accepted}

  \abstract
   {Large-scale astronomical surveys from ground-based as well as space-borne facilities have always posed significant challenges concerning the problem of automatic extraction and flux estimate of sources. The recent explosion of surveys  in the mid-and far infrared, as well as in the sub-millimeter, brings an increase to the complexity of the source extraction and photometry task because of the extraordinary level of foreground/background due to the thermal emission of cosmic cold dust. The maximum complexity is likely reached in star forming regions and on the Galactic Plane, where the emission from cold dust is dominant.}
   {We present a new method for detecting and measuring compact sources in conditions of intense, and highly variable, fore/background.}
   {While all most commonly used packages carry out the source detection over the signal image, our proposed method builds from the measured image a "curvature" image by double-differentiation in four different directions. In this way point-like as well as resolved, yet relatively compact, objects are easily revealed while the slower varying fore/background is greatly diminished. Candidate sources are then identified by looking for pixels where the curvature exceeds, in absolute terms, a given threshold; the methodology easily allows us to pinpoint breakpoints in the source brightness profile and then derive reliable guesses for the sources extent. 
Identified peaks are fit with 2D elliptical Gaussians plus an underlying planar inclined plateau, with mild constraints on size and orientation. Mutually contaminating sources are fit with multiple Gaussians simultaneously using flexible constraints.}
   {We ran our method on simulated large-scale fields with 1000 sources of different peak flux overlaid on a realistic realization of diffuse background. We find detection rates in excess of 90\% for sources with peak fluxes above the 3$\sigma$ signal noise limit; for about 80\% of the sources the recovered peak fluxes are within 30\% of their input values.}
   {}


	\authorrunning{Molinari et al.}
	\titlerunning{Source extraction and photometry with complex backgrounds}
   \maketitle
%

\section{Introduction}
\label{intro}

Automatic source detection and photometry is a long standing problem in many areas of modern astrophysics. A need first felt when trying to automatically extract information from large-scale photographic plates, it is now an astronomer's everyday activity in the era of large-format detector arrays and fast-mapping facilities virtually at all wavelengths. 
The very fact that new packages and novel approaches keep being developed, however, is an indication that the problem does not have a single solution; this is mostly due to the very different working conditions we have to face depending on the source properties to be extracted and the characteristics of the background on top of which sources are found. In other words, the extraction of very crowded stellar point-like optical sources in relatively low background presents very different challenges from the extraction of very faint objects in far-infrared deep cosmological surveys; which is again different from detecting and measuring variable-size compact sources in intense and complex backgrounds. It is then not easy, and certainly beyond the scope of this paper, to provide a comprehensive review and benchmarking of the major approaches that have been proposed over time, or discuss their advantages and shortcomings, since different approaches are generally optimized to the different conditions found in an astronomical image. 
 
\begin{figure}[b]
\resizebox{\hsize}{!}{\includegraphics{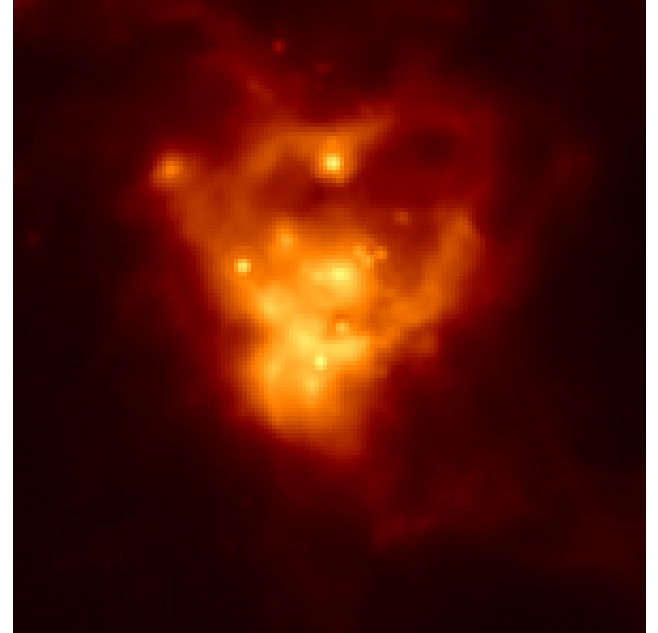}}
\caption{Spitzer/MIPS 24\um\ image of IRAS23385+6053 (Mol160). Sources of all sizes are visible from pure point-like to resolved blobs which however the eye immediately detect as separated features from the more distributed background.}
\label{image_example}
\end{figure}

The first step in source finding always implies thresholding. It can be done directly on the measured map as in DAOFIND \citep{stetson87} or SExtractor \citep{bertin96}. These two methods have seen a wide application to the analysis of fields where the background is relatively well behaved. \textit{Spitzer} imaging surveys, however, have shown that such conditions are hardly found in the mid-infrared and far-infrared toward regions like star forming clouds or the inner Galactic Plane, (e.g. \citealt{car09}, \citealt{rebull07}). Convolution is used in DAOFIND or SExtractor to enhance desired features but that requires that a typical source "template" can be defined. Even then, finding peaks requires that a fixed threshold is set, and that is hardly possible when the background greatly varies. SExtractor's ability to estimate a variable background for thresholding is put to a hard test in presence, like in Fig. \ref{image_example} (a typical high-mass star formation region images at 24\um\ \citep{mol08b} of compact sources of various sizes, from purely point-like to slightly resolved, distributed with a certain degree of crowding on top of a diffuse back/foreground which present structures at all scales. 

In other approaches, nonlinear matched filtering has been used, as in MOPEX \citep{makovoz05}, to build a point-source probability figure on which to perform the thresholding. Similarly using a source template as a "prior" knowledge, \citet{savoliv07} used the Bayesian Information Criterion as a figure of merit. Also in these cases however, these approaches may be underperforming. 

It is true, however, that an experienced eye has little trouble in recognizing where sources are, what is their size and where they give way to the background irrespectively of the absolute level of the source or background signal. This is because the eye reacts to \textit{changes} in signal rather than to absolute values. Can we reproduce this behavior in a computer program without having to resort to filtering the image with any sort of source template ?

The idea we want to propose here is to let the computer deal with the \textit{curvature} of the brightness distribution in the image, rather than with the direct intensity image, not only to detect sources but also to give a reliable estimate of their size. Recent studies for the characterization of cosmological filaments \citep{bond10} pursue a similar line of thought.

Estimating the flux of an identified source is the subsequent step. Again, in the particularly difficult conditions of crowding and variable size sources embedded in variable background regions, the relatively standard methods of aperture photometry or "fixed-template" source fitting (e.g. the PSF) may show their limits. We attempt a more relaxed approach by fitting variable-size Gaussians to identified peaks; we limit the impact of increasing the fit free parameters by a careful estimate of reliable initial guesses for the fit parameters and using flexible fitting engines which allow us to better control the quality of the convergence compared to a completely unconstrained fit. 

In the following sections we will illustrate in detail the proposed detection and photometry methods, which we called CUTEX (CUrvature Thresholding EXtractor). We will also present in a synthetic form the results of extensive testing that was done on simulated images. We plan extensive applications of this  code to the source extraction from the large scale Galactic photometric surveys that the \textit{Herschel} satellite will carry out. The code is written in IDL, and while we plan to make it freely available in due time, it can be obtained for testing by contacting directly the authors.

\section{Curvature-based source detection}
\label{detection}

The method consists of a two-step process. We first build a set of images containing the 2$^{nd}$-order differentiation of the signal image in four different directions; the "curvature" images obtained are then thresholded to mask pixels where the curvature suggests the presence of a compact object. This pixel mask is then analyzed to group together adjacent pixels to positively identify a candidate source. Once groups of masked pixels are identified, the curvature in each group is analyzed to verify if statistically significant local peaks are present which lead to a segmentation of the group of pixel in individual candidate sources. We explain all the steps in more detail in the subsequent paragraphs.

\subsection{The Lagrangian differentiation}

The input image is differentiated along 4 preferential directions: X, Y and the two diagonals. Since the image is a set of discrete pixels with a constant spacing the differentiation is performed using the Lagrangian differentiation method. The Lagrangian formulas (for numerical interpolation, integration and differentiation) are expressed explicitly in terms of the dependent variable involved, rather then in term of its finite differences. 

Let's assume we have a function whose values are known over a set of discrete points, $f(x_k)$ for $x_k, k=0,n$. At any given point the function is approximated as  

\begin{equation}
\label{lagap}
  f(x) = \sum_{k=0}^{n} l_{k}(x)f(x_{k}) + E(x)
\end{equation} 

where $E(x)$ is a correction term which is generally a 2$^{nd}$-order infinitesimal and will be neglected, and $l_{k}(x)$ are the \textit{Lagrangian Coefficient functions}. The derivative of $f(x)$ for any $x_i$ of the discrete set of points is then

\begin{equation}
  f'(x_{i}) = \sum_{k=0}^{n} l'_{k}(x_{i})f(x_{k})
\end{equation}

\begin{figure}[h]
\resizebox{\hsize}{!}{\includegraphics{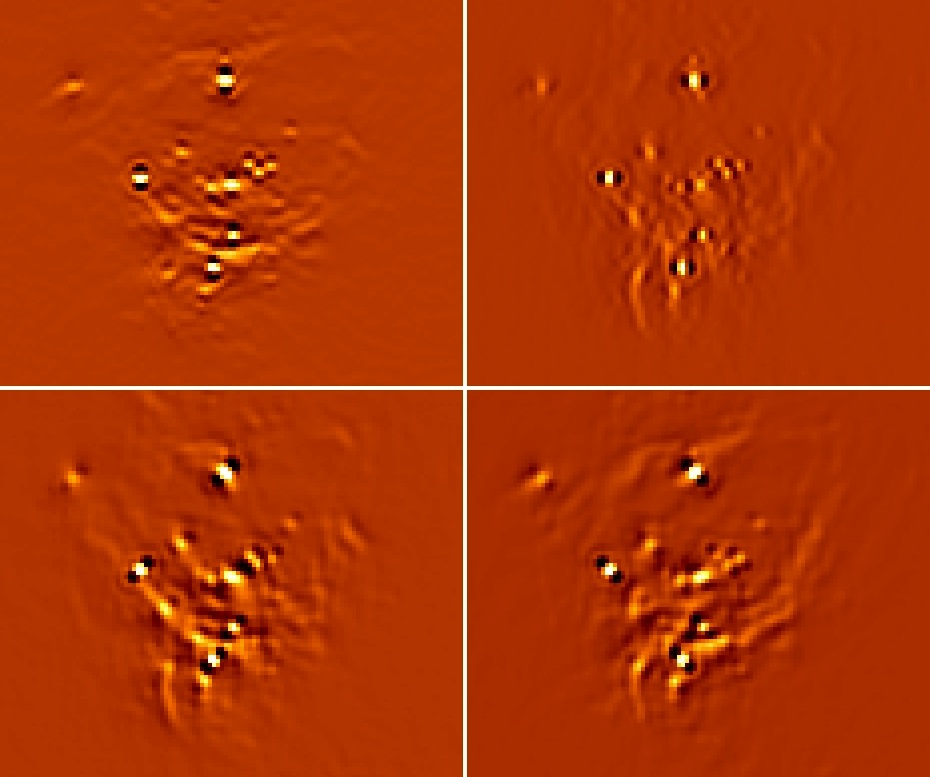}}
\caption{$2^{nd}$-order derivatives images of the same field of Fig. \ref{image_example}; derivatives have been computed along columns (top-left), rows (top-right) and the two diagonals (bottom panels). As a matter of convention, we decided to invert sign on the derivatives (which should be negative on the source peaks) for purely cosmetic purposes.}
\label{der2_example}
\end{figure}

The $l'_k(x)$ can be computed (see \citealt{hild56} for the full derivation), but in the case of equally spaced $x_k$, the  Lagrangian coefficient functions have been tabulated rather extensively for various values of $n$. An important choice to be made is how many points should be considered in the differentiation. The Lagrangian formulas are often implemented in their 3-points form, i.e. $x_k$ with $k=0,2$. In case of real astronomical images where the noise is a non-negligible component, however, the 3-points differentiation can be too sensitive to glitches and spikes which could mimic a point source. Assuming that the pixel grid in an image is such that a pixel is 1/3 of the beam FWHM, we found after extensive testing that a 5-points derivative is more adequate to describe the curvature of a compact source, being at the same time more robust against noise spikes and glitches. The formula for the value of the 5-point 1$^{st}$-derivative at any pixel in the map (e.g. pixel 0) is then expressed as a function of the values of the map itself in the pixels -2, -1, 1 and 2 along the chosen direction as:

\begin{equation}
   f'_{0}=\frac{1}{12d}\left({f_{-2}-8f_{-1}+8f_{1}-f_{2}}\right) \\
\end{equation}

where $d$ is the constant interval between the discrete points ($d=1$ as we use all pixels in rows, columns and diagonals). Running the differentiation twice will yield the image of the 2$^{nd}$-order derivatives; for brevity in the following we will use the symbol \dder\ for the 2$^{nd}$-order derivatives processing. Fig. \ref{der2_example} shows the result of applying the above treatment on the image of Fig. \ref{image_example}. 

As a matter of convention, we decided to invert sign on the derivatives (which should be negative on the source peaks) for purely cosmetic purposes. The images show that the diffuse emission which is giving us so much troubles for the source detection  in the signal image has been removed. Sources are clearly standing out so that the source detection can now be easily attained with one single threshold. It can be noted that complex filamentary structures are visible in the various panels of Fig. \ref{der2_example}; they survive the double differentiation since they are high spatial frequency structures. However, it is important to point out that the structure appearance changes depending on the differentiation direction; by requiring that the curvature threshold is exceeded for all four differentiation direction we basically remove the chance that such structures are extracted as compact objects. 


\subsection{Extraction of sources}

The \dder\ images are used to determine the location and properties of the sources. A mask image is generated, containing the pixels where a curvature threshold $\zeta _{th}$ is exceeded for all differentiation directions. These curvature mask pixels (CMP) are then grouped together if they are contiguous in space to produce what we will call $clusters$ which will contain a variable $n_{CMP} \geq1 $ number of CMPs. The CMP clusters are the basis to establish the existence and location of sources in the original signal image. It is clear that for any given source the lower the threshold used, the higher number of CMPs will be found in the corresponding CMP cluster. We express the curvature detection threshold in units of \sigdder, the r.m.s. of the \dder\ image (using, for the detection in each \dder\ image, the \sigdder\ determined on that \dder\ image). We ran a series of simple tests to characterize the relationship between $\zeta _{th}$ and $n_{CMP}$ (see \S\ref{cmp_members}), and determined that for thresholds $\zeta _{th} \geq 0.5$\sigdder\ (lower thresholds would lead to excessive false detections) the minimum $\tilde n_{CMP}$ for a reliable source detection is 3.

The position of the tentative source or sources within a CMP cluster is determined at the pixel(s) $p_i$ in the cluster which is/are statistically significant local maxima in the average of the 4 \dder\ images. In particular we positively identify a candidate source location if its curvature value is at least 1\sigdder\ above the curvature value of all the surrounding CMPs. Extensive simulations in various source crowding configurations showed us that values higher than 1\sigdder\  tend to miss close companions. Lower values ensure de-blending of objects as close as 1 PSF but increase the number of false positives, so that the total "error rate" of the detection algorithm does not substantially depend on the exact choice. If no statistically significant (in the above sense) local maximum $p_i$ is found, then the position of the source is taken as the average of the coordinates of all CMPs in the cluster. We stress again that the detection threshold is not on the signal level in the image (and indeed a 0.5$\sigma$ detection threshold would be unreasonably low), but on the level of curvature in the \dder\ image. We will discuss in \S\ref{size_guesses} how the \dder\ images can also be used to obtain reliable first guesses of the source size.

\section{Source Photometry with Gaussian fitting}

This method is primarily being developed for image analysis of far-infrared and sub-millimeter continuum images which will be acquired with the PACS and SPIRE cameras on board the Herschel satellite. At these wavelengths 
the emission comes from cold dust mostly organized in diffuse ISM structures with complex morphology at all scales  and from clumps and cores associated with star formation, sources which are dominated by extended envelopes which may be resolved spatially. Even stellar objects like Post-AGB are surrounded by extended envelopes which may not be point-like to the beam. These conditions are hardly optimal for aperture photometry because the background varies to such an extent that the usual estimate in a surrounding annulus may not be at all an accurate estimate for the background on the source position. In addition, clumps and cores of cold dust come in all sizes, and a single aperture to be applied for source photometry over a large sky area (e.g. a star forming region, or a patch of the Galactic Plane), which would be highly desirable for homogeneity purposes, can hardly be found; last but not least, the source crowding of variably extended sources makes aperture photometry even more difficult. On the other hand the classical alternative of PSF-fitting is also difficult to use because sources are not uniformly point-like across an image, like in a stellar field in the optical or near-infrared, and a fit with a fixed shape would deliver highly inaccurate results. 

To estimate source photometry we then make the assumption that the source brightness distribution can be approximated with a two-dimensional Gaussian profile with variable parameters, including gaussian size and orientation. However, since the sources we want to extract and measure may sit upon an intense and complex background spatially varying on all scales, we will model each source with an additional planar plateau at variable inclination and inclination direction which we want to simultaneously fit to remove background emission. 

The function we will fit is then of the form:

\begin{multline}
F = F_0 \cdot exp \biggl( -\frac{1}{2} \biggl\{ \biggl[ \frac{(x-x_0) \cos{\theta} + (y-y_0)\sin{\theta}}{\sigma _x }  \biggr] ^2 + \\ 
\biggl[ \frac{(y-y_0) \cos{\theta} - (x-x_0)\sin{\theta}}{\sigma _y }   \biggr] ^2 \biggr\} \biggr)  + [ b_0 + (x-x_0)\, b_1 + (y-y_0)\, b_2 ]
\label{gaussian}
\end{multline}

While the concept is certainly not new (e.g. GAUSSCLUMPS, \citet{SG90}), obtaining an accurate and credible result from the fitting is critically dependent on the source extent, the presence of other nearby sources and the intensity and morphological complexity of the background. In \citet{mol08} we used the Gaussian fitting approach available in the AIPS package to derive sub-millimeter and millimeter fluxes toward massive YSO envelopes; in order to have credible results, however, we were forced to examine the fields one by one and set constraints to some of the fitting variables by estimating them independently. Although very time-consuming, the procedure was necessary in the cases where the analysis of the brightness profiles clearly established the presence of a plateau underlying the core; in these cases packages like CLUMPFIND \citep{williams94} will integrate the flux down to a saddle point or a given threshold value, often overestimating the flux. For example the subsample of YSO fields which we analyzed in \citep{mol08} was extracted from a larger sample that was also analyzed using CLUMPFIND in \citet{bel06}. The gauss+plateau approach, which was applied manually at the time, delivered fluxes which were a fraction between 30\% and 100\% of the CLUMPFIND ones. 

Here we would like to ease the procedure of source fitting by trying on one side to automate a reliable guess for the initial fitting parameters, and on the other to set clever fitting constraints.

\subsection{Initial guess for source sizes}

\label{size_guesses}

Regarding the source size parameters (FWHM in two direction plus position angle), the  \dder\ images contain information that can be used to extract an initial guess on the size of the candidate sources. Before the source merges into the underlying plateau the brightness profile will change from convex to concave curvature; the second derivatives will change sign and reach a minimum  (remember our "cosmetic" convention to invert the sign of the second derivatives to make them positive peaks) before joining the regime characteristic of the curvature fluctuations of the extended emission. Assuming a Gaussian source profile, the position of this local minimum can be computed analytically and it is easy to verify that its distance from the Gaussian peak is always a factor of 1.47 larger than the point where the Gaussian has its half-maximum value.

We then analyze the four \dder\ images by following the profile from each curvature peak and determining the pixels (relative to the source positions) on each side where the \dder\ image values reach a local minimum (always keep in mind the inverted sign). This is done on each side of the source along the four differentiation directions, using each time the proper \dder\ image, and results in  a set of 8 points to which we will fit an ellipse with variable semi-axis and position angle $\theta$. Clearly in a non-ideal environment with source confusion, variable extended emission and noise it may happen that in a certain direction a minimum is not reached reasonably close to the objects and is hence unreliable. Along each direction then, the minimum \dder\ points are considered valid if their distances from the source peak are within 20\% of each other, otherwise only the nearest point is kept. The two semi-axis of the fit ellipse are then divided by 1.47 (see above) and multiplied by two to obtain the FWHM. If the latter is found larger than three times the PSF we will flag the size initial guess as unreliable and put it equal to the PSF. The FWHM guesses are then divided by 2.354 and provided as initial values for $\sigma _x, \sigma _y$ in the 2D-Gaussian fit (see Eq. \ref{gaussian}).

\subsection{Identifying groups of potentially blended sources}
\label{source_groups}

A fit for a source involves 6 parameters for the Gaussian ($F_0, x_0, y_0, \sigma _x, \sigma _y,$ and the position angle $\theta$) and 3 parameters for the plateau ($b_0,b_1$ and $b_2$), and an entirely free least-squares fit may easily converge to unsatisfactory solutions when compared with what the eyes see. Setting constraints on certain fitting parameters is a viable solution for a limited number of sources, but is clearly impractical in case of large-area surveys. The entire problem is of course complicated in case of blended sources, as it is often the case in star forming regions; a possibility is to  fit individual sources and make an estimate of their contamination to nearby ones and iteratively do this for all sources which are close enough to suffer from reciprocal contamination until the estimated parameters do not appreciably change any more. We instead chose the conceptually easier approach to fit {\it simultaneously} all sources which may be potentially mutually contaminating one another. 

\begin{figure}[h]
\resizebox{\hsize}{!}{\includegraphics{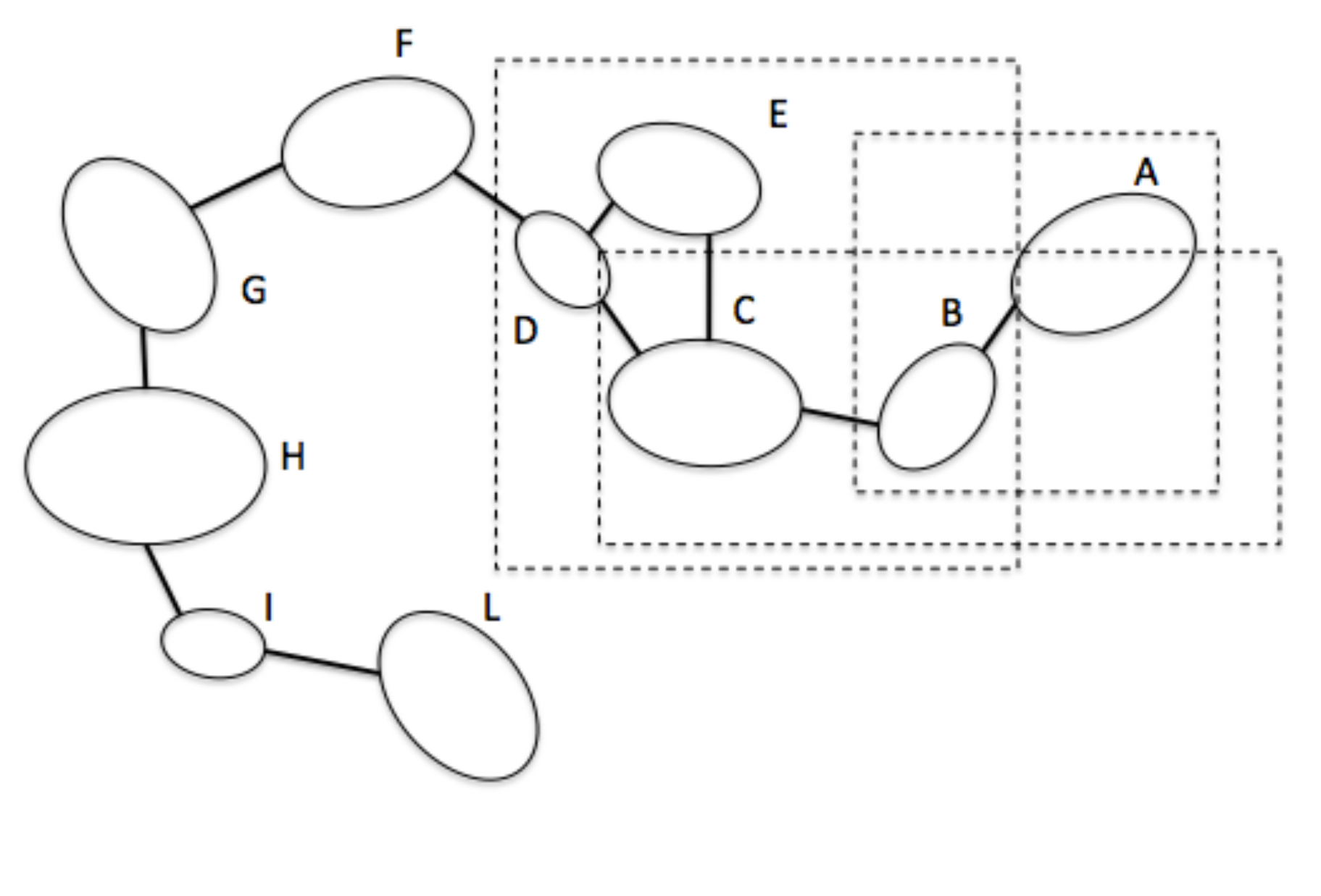}}
\caption{Diagram showing a typical source grouping run. Sources A and L will end up in the same group although their reciprocal distance is larger than the 2$\times$FWHM adopted threshold.}
\label{fof}
\end{figure}

The detection file produced by CUTEX is parsed and each source is searched for neighbors located within a given threshold radius (set to twice the PSF as a default, but see \S\ref{tune_thresh}). Ideally, we should follow a "friends-of-friends" approach to make sure that each source gets a proper accounting of potential contamination from its closest neighbors; Fig. \ref{fof} shows how a typical grouping run would end up with quite a large group. The simultaneous fit of such a group is impractical as all sources should be fit with same background model and simulations carried out using spatially complex backgrounds (see \S\ref{simul}) show us that source fluxes are not satisfactorily recovered. The fit to a source is then performed simultaneously with its nearest neighbors \textit{only} (within the adopted distance threshold). For example, the fit to source A requires that also B is simultaneously fit; while, however, the so-estimated flux of source B will be sufficiently accurate to account for its contamination to source A, its best flux estimate will require that B is fit simultaneously with A and C, too. In turn, the accurate flux estimate of source C requires simultaneous fitting with B, D and E. In other words, we will fit the $i_{th}$ source simultaneously with its subgroups of nearest neighbors and each time recording the best fit parameters only for the $i_{th}$ source.

\subsection{Constrained fitting}

When multiple Gaussians have to fitted together, the number of parameters which can vary at any given time becomes high enough that extra measures have to be taken to hope for meaningful convergence. The fitting engine provided by the MPFIT package \citep{mark09} proved ideal for this task. MPFIT is a general fitting package where any user-provided analytical formulation can be defined and fit to a set of data. MPFIT's real value-added, however, is the way in which constraints to variables can be provided. Apart from the standard possibility to keep any of the parameters fixed in the fit, MPFIT allows for any parameter to be limited between a lower and/or an upper limit; an even more interesting feature is the ability to constrain any parameter to vary "tied" to any other. In other words, we tie the X$_i$,Y$_i$ peak positions of $n-1$ sources to the position X$_1$,Y$_1$ of the first source in the group. In this way we allow  the source positions to adjust not individually but {\it as a group}, and tests have shown that this significantly improves convergence toward credible solutions. 

Another handy feature provided by MPFIT is the ability to have fitting parameters free to vary within boundaries. we used this feature to have the source size values vary in the fit by 20\% at most from their initial guesses (\S\ref{size_guesses}) to allow finer adjustments on the real signal image; in case the initial guesses were flagged as unreliable, the size was left a completely free parameter. This assumption will be further discussed in \S\ref{size_assum}.

The Gaussian fitting for single source fitting is carried out over a portion of the image whose size can be chosen by the user, and that we set by default to 4 times the PSF. This choice generally ensures a sufficiently large fitting area to allow the algorithm to perform reliable estimate of the local background; a smaller area would be almost entirely dominated by the source emission and the background fit would be poorly constrained. In case more sources are to be simultaneously fit, the image portion over which the fitting is carried out will be the minimum box that contains all sources positions, increased of the amount of a PSF on each side. The initial guess of the background in the fitting box is done by first masking all pixels where candidate sources are located (the masked area is twice the source size estimated as described in \S\ref{size_guesses}) and taking the median of what's left in the fitting window.


\subsection{Fit results}

Sources are fit with a 2D Gaussian function (or with $n$ Gaussians in case of source groups, see \S\ref{source_groups})  plus a planar plateau. A noise map can be provided for proper least-squares fitting to derive meaningful $\chi ^2$ fit results; if this is not available the fit can be done with uniform weights, but in this case the uncertainties will be upper limits.

\begin{figure}[t]
\resizebox{\hsize}{!}{\includegraphics{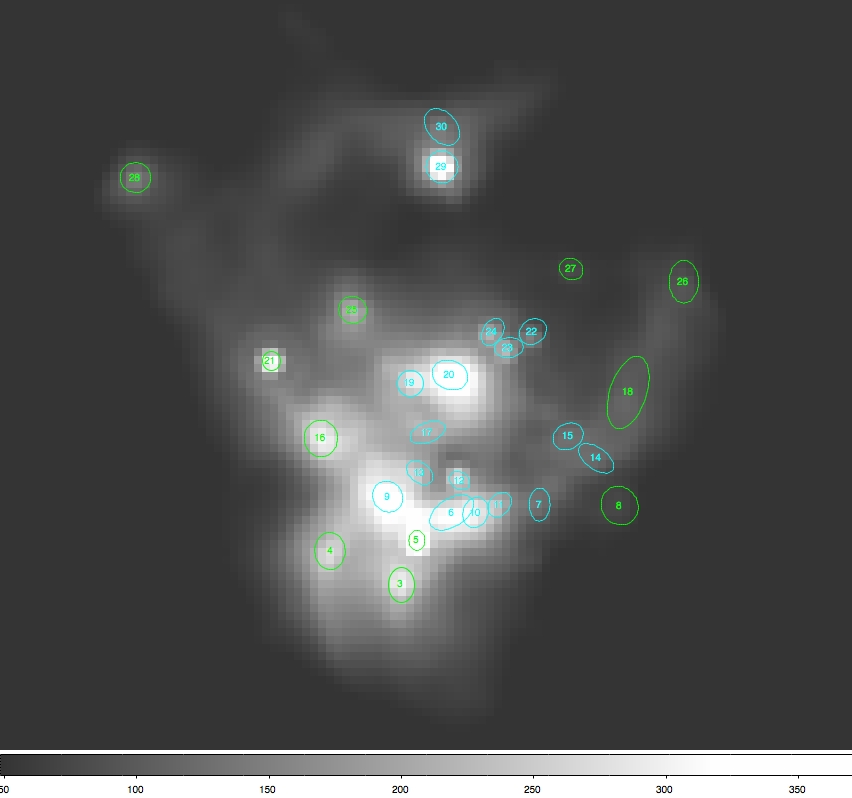}}
\caption{Results of the detection and photometry procedure over the same Spitzer/MIPS 24\um\ field of Fig. \ref{image_example}. Extracted sources are reported as ellipses whose size represents the FWHM and orientation of detected compact structures. Green ellipses represent sources fit individually, while blue ellipses are for sources which were fit in multiple groups.}
\label{results_example}
\end{figure}

The primary output product is a photometry file which includes essential information on the position of each fit source, both in pixel and in equatorial coordinates, the source size in X and Y and the position angle measured CCW in degrees, the peak as well as the integrated flux, the uncertainties for all parameters and the estimate of the background at the position of the source. The integrated flux is obtained as 

\begin{equation}
\label{intflux}
S_{int}= S_0 2\pi\ \sigma _x \sigma _y /\Omega _b
\end{equation}

 where the the beam solid angle is 
20
\begin{equation}
\label{beam}
\Omega _b = \frac{2\pi}{8 ln2} \theta _b ^2
\end{equation}

and $\theta _b$ is the beam FWHM.

The covariance matrix provided by the MPFIT fitting engine provides the uncertainties on all fit parameters. Formal uncertainties on derived quantities like the integrated flux are obtained via standard error propagation.
The output file is in ASCII tabular format and can be easily imported into most used applications for image visualization (e.g.  Aladin). The code also create files in the format ({\it region} files) ready to be read in from application like SAOImage to overlay the size and orientation of the fit 2D gaussians on the displayed image. An example is presented in Fig. \ref{results_example} in which all compact sources visible by eye are recovered with one run of the code. To obtain a similar result using, e.g., DAOFIND, we should have carried out multiple runs with different source FWHMs and detection thresholds.

\begin{figure}[t]
\resizebox{\hsize}{!}{\includegraphics{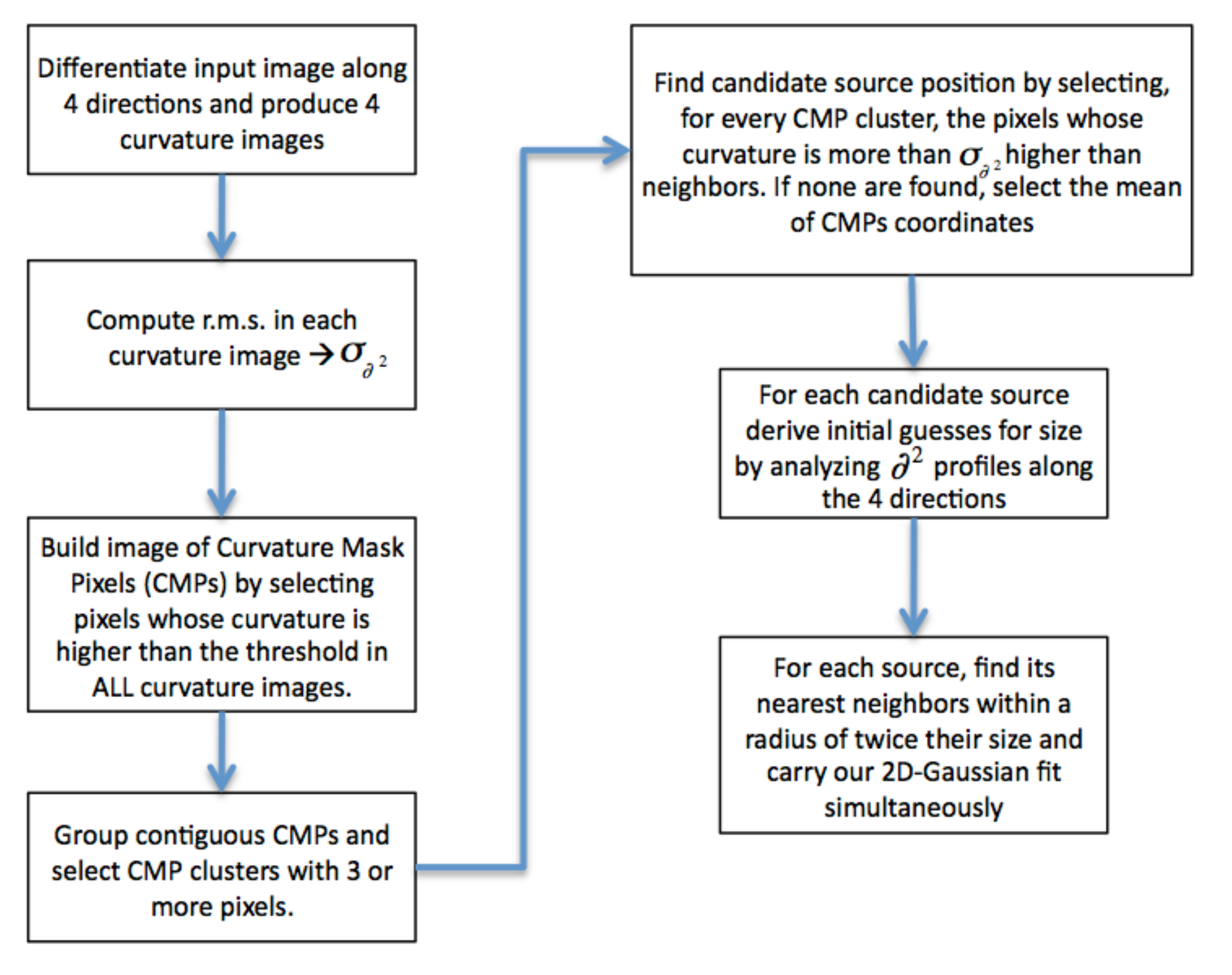}}
\caption{Flow chart for the extraction and photometry method.}
\label{flowchart}
\end{figure}

Finally, it is easy from the photometry file to generate source-subtracted images to verify the accuracy of the fit sources allowing a statistical characterization of the residuals.  In this way we can determine a perhaps more realistic figure to characterize the significance of the extracted sources, by dividing the source peak flux by the $r.m.s.$ of the local residuals after the fit source+background has been subtracted. In Fig. \ref{flowchart} we summarize the logical flow of operations in our extraction and photometry approach.

\section{Code Testing}

We tested the code performances doing runs at different detection thresholds on simulated fields containing four different populations of synthetic sources at different peak flux and source size levels. The test sources will therefore exhibit a range of integrated fluxes. The reliability of the method will be measured comparing the recovered fluxes with respect to the "true" fluxes we used to generate the synthetic images. A diffuse emission component which should be representative of the diffuse emission toward the Galactic Plane in the far-infrared is also added; we also include a white noise component representing the expected detector noise for the Herschel imaging cameras. We detail the process in the following paragraphs.

\subsection{Construction of simulated images}
\label{simul}

\begin{figure*}[t]
\resizebox{\hsize}{!}{\includegraphics{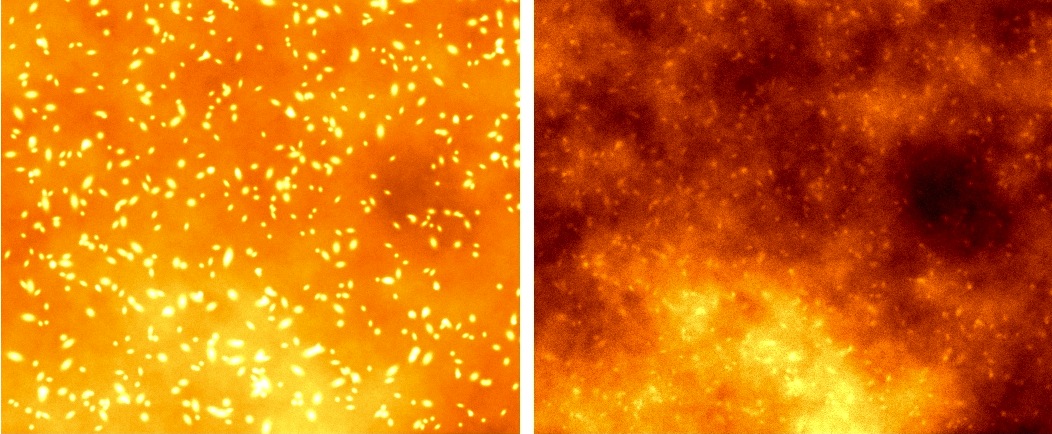}}
\caption{Simulated fields at 250\um. The angular size is 1\adeg\ and the pixels size is 6\asec. The case for source peak fluxes of 1 Jy/pxl and 0.1 Jy/pxl are reported in the left and right panels, respectively. The color stretch in the two figures is different to put emphasis on the sources, but the level of the diffuse emission in the two fields is the same.}
\label{sim_fields}
\end{figure*}

A 1\adeg x1\adeg, field was populated with 1000 synthetic sources characterized by a 2D Gaussian spatial brightness distribution with the intent to mimic how compact dust condensations (enveloped around YSOs or dense molecular cores or clumps) would appear in the far-infrared or sub-millimeter; the maps are indicatively produced for a wavelength of 250\um, that is one of the photometric bands of the SPIRE camera \citep{gri09} on-board the Herschel satellite. Sources positions were randomly extracted using a uniform distribution probability. All sources were assumed to have the same peak flux with source Full Widths at Half-Maximum (FWHMs)  randomly assigned using a uniform probability distribution in an interval between 18\asec\ and 36\asec, corresponding to one time and twice the size of the Point Spread Function (PSF) of the Herschel Telescope at 250\um. In addition, the sources were allowed an elliptical shape with variable axis ratio up to a maximum of 2. 

The simulated field were summed to an extended emission component generated using a so-called "fractional Brownian motion" image \citep{stutzki98}. This class of images is characterized by two properties of their Fourier transform: (i) the power spectrum has a power-law behavior as $|F(k)| ^2 \propto k^{-\beta}$, and (ii) the distribution of the phases is completely random. Images are generated in Fourier space and transforming back to the direct space. Given a phase distribution, the $\beta$ exponent determines the noise level of the image: $\beta=0$ corresponds to the white noise case, and an increase of $\beta$ produce a smoother image. We used a realization using $\beta=3.4$ which is empirically found representative of the fluffy and filamentary structure of the ISM in the far infrared as appearing in the first images obtained with Herschel \citep{moli10c}. The flux density of the simulated diffuse emission component varies between 100 and 1000 MJy/sr which is comparable to the range of fluxes reported in the 100\um\ IRAS maps of the Galactic Plane at longitudes between $l$=30\adeg\ and $l$=40\adeg.

White noise was injected doing a random realization using a normal distribution with a standard deviation of 20 mJy/pxl, well matched to the predicted noise levels in the Galactic Plane maps from the Hi-GAL Key-Project \citep{moli10a}. The noise level on the simulated maps are, however, higher due to the contribution of the fluctuations by the extended emission. The r.m.s. measurement done on several places of the map of the diffuse emission+noise yields values from 30 to 60 mJy/pxl.

Four simulated fields were generated where the peak flux of the synthetic sources was 0.1, 0.2, 0,3 and 1 Jy/pxl; the pixels size in the simulated fields is 6\asec, or 1/3 of the Herschel PSF size at 250\um. The properties of the diffuse emission and the injected noise levels were the same in the four fields, as well as the location of the 1000 sources. Fig. \ref{sim_fields} shows the simulated field for source peak fluxes of 1Jy/pxl and 0.1 Jy/pxl. Given the measured r.m.s. levels on the map of the simulated diffuse emission+noise (see above), the peak flux  levels of the injected sources would correspond to average signal standard detection thresholds between 2.5 and 25$\sigma$.

\subsection{Test Results}

\begin{figure}[t]
\resizebox{\hsize}{!}{\includegraphics{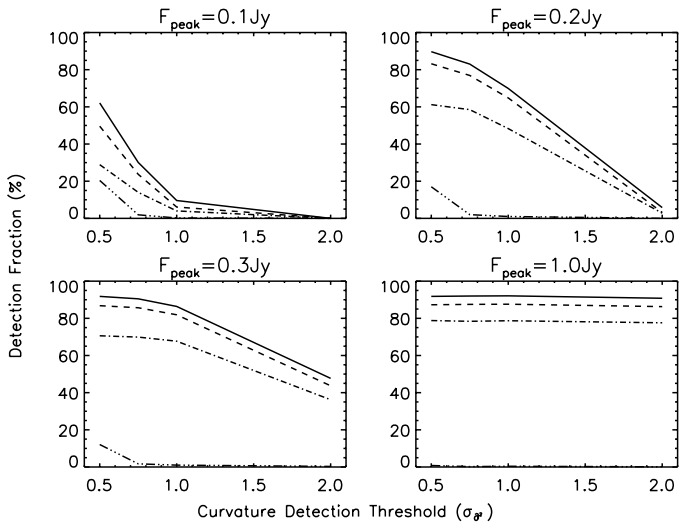}}
\caption{Fraction of recovered sources as a function of the curvature detected threshold. The various panels report the results for the different sources peak flux used as displayed. The full line is used for the total recovery fraction; the dashed and the dash-dotted lines are the fraction of sources for which the recovered peak flux and integrated flux, respectively, are  within 30\% of the input value. The triple-dot-dashed line represents the fraction of false positives over the number of input stars.}
\label{det_results}
\end{figure}

Four CUTEX runs with four different detection thresholds were carried out on each of the four simulated fields. The four detection thresholds adopted are $\zeta_{th}$=0.5, 0.75, 1.0 and 2.0\sigdder. We remind again the reader that the detection threshold is applied on the curvature images; the value of \sigdder\ is determined separately for each \dder\ image (i.e., for each differentiation direction), and the threshold check is applied on each \dder\ image independently using the proper \sigdder. The output source catalogues were then matched to the input ''truth tables" adopting a matching radius of 2 pixels. Fig. \ref{det_results} reports the detection statistics as a function of the detection threshold for each assumed peak flux (different colors). The figure shows the percentage of recovered sources (full line) and the percentage for which the recovered peak fluxes (dashed line) and integrated fluxes (dash-dotted lines) are within 30\%\ of the input value.

The results in terms of detection are quite encouraging and show that we recover more than 90\% of the input sources for peak fluxes of 0.2 Jy/pxl (corresponding to about 5$\sigma$ r.m.s.) with the peak flux within 30\% of its input value in more than 80\% of the input sources. The situation is a bit worse for the fainter objects (corresponding to 2.5$\sigma$ sources), where this fraction drops to 60\%.

\begin{figure}[t]
\resizebox{\hsize}{!}{\includegraphics{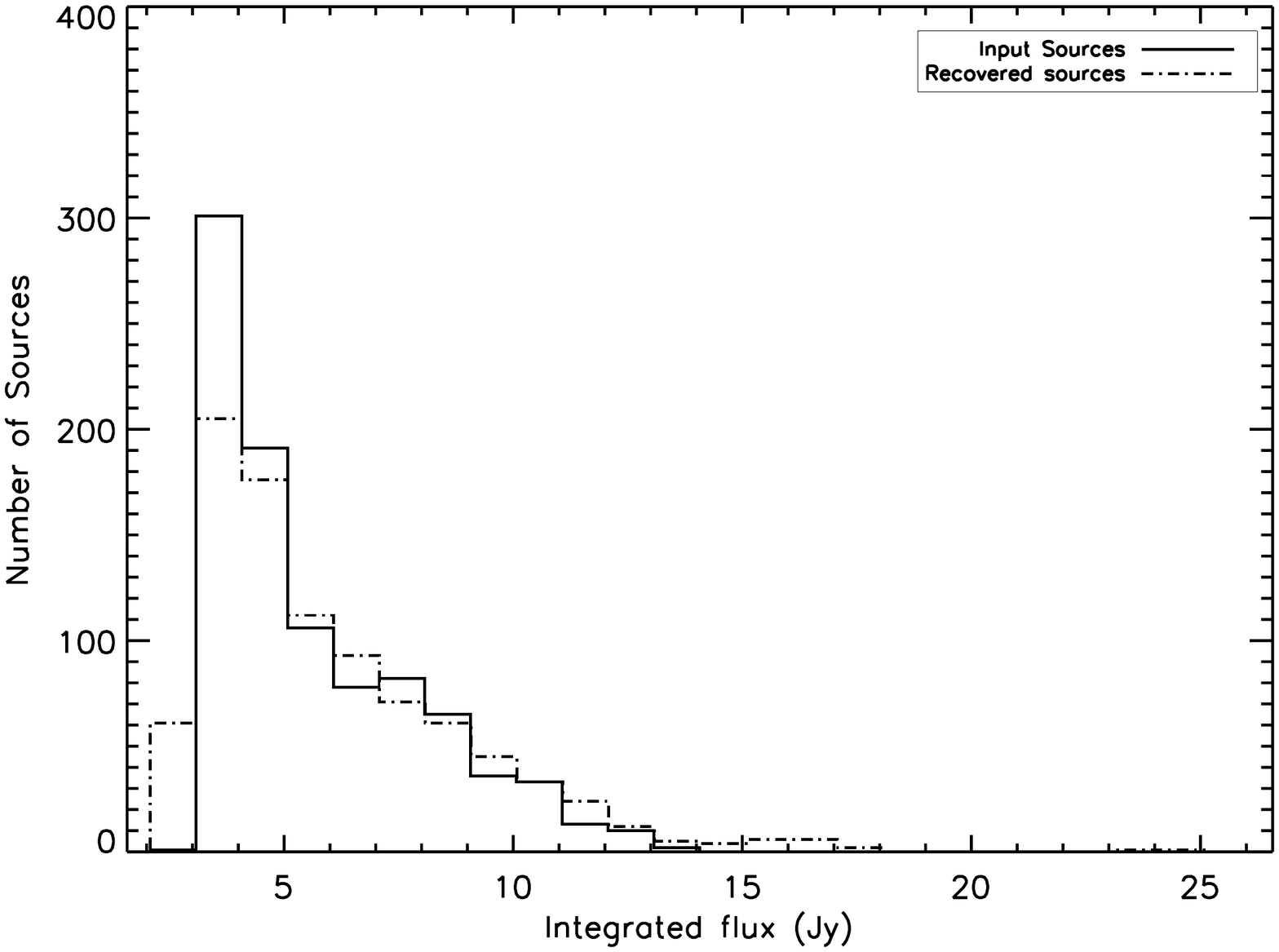}}
\resizebox{\hsize}{!}{\includegraphics{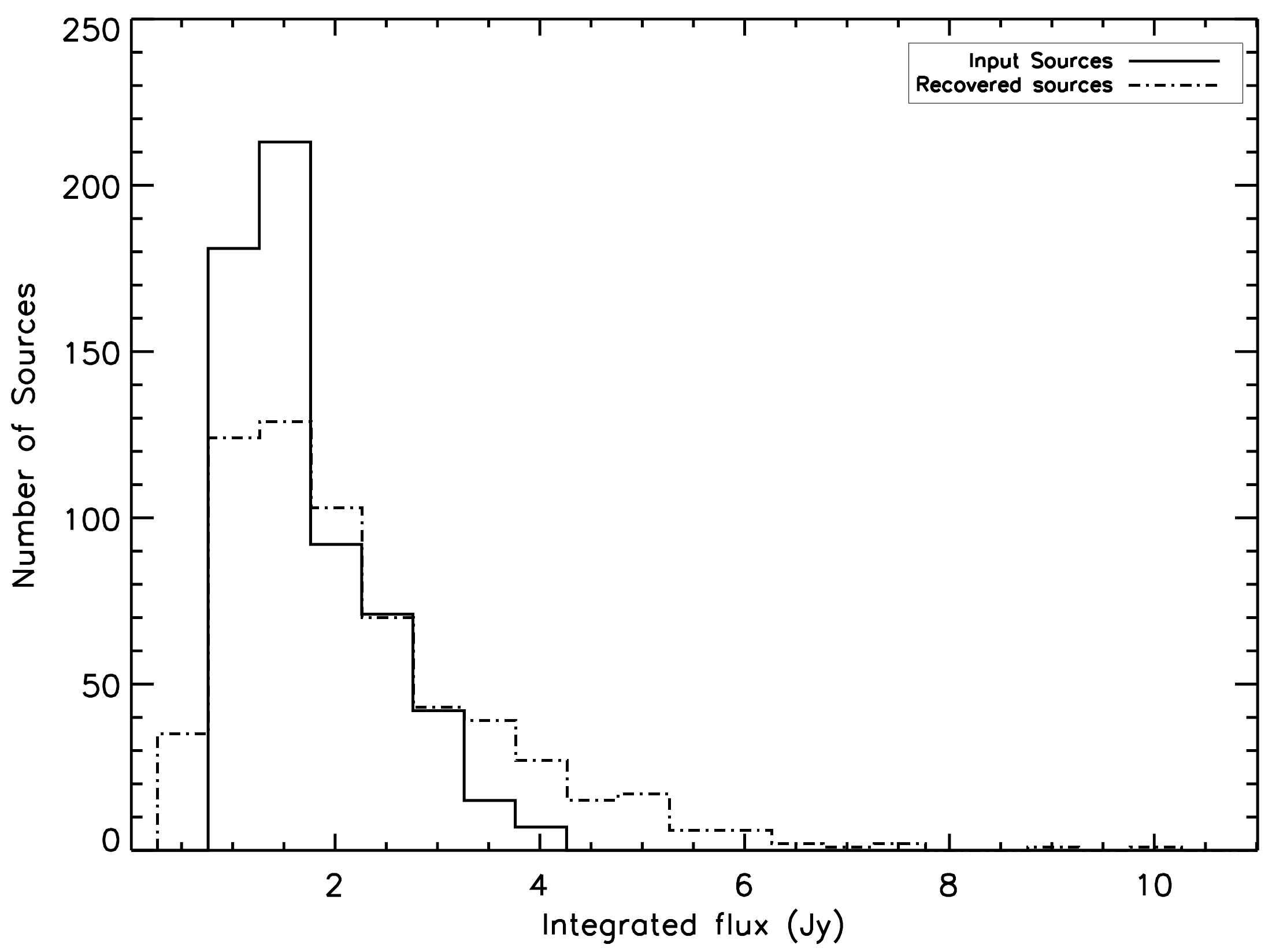}}
\caption{Histograms for the distributions of integrated fluxes for the sources which were positively recovered in the simulations for F$_{peak}$=0.3 Jy/pxl sources (top panel) and for the F$_{peak}$=0.1 Jy/pxl sources (bottom panel). In each figure the full line represent the distribution of input fluxes, while the dot-dashed line is the distribution of recovered fluxes. The results reported are those for a curvature threshold of 0.5\sigdder.}
\label{int_flux}
\end{figure}

The recovered integrated fluxes show a larger scatter with respect to the peak fluxes (compare for each color in Fig. \ref{det_results} the dot-dashed and the dashed lines), but the recovered and input integrated fluxes are overall in good agreement. Fig. \ref{int_flux} shows (for the cases of 0.3Jy/pxl and 0.1Jy/pxl peak flux sources) that the flux distributions are essentially coincident. In the F$_{peak}$=0.1 Jy/pxl simulation the situation is worse as the larger sources tend to show less contrast with respect to the underlying diffuse emission and are therefore fit as larger objects than they intrinsically are.

\begin{figure}[t]
\resizebox{\hsize}{!}{\includegraphics{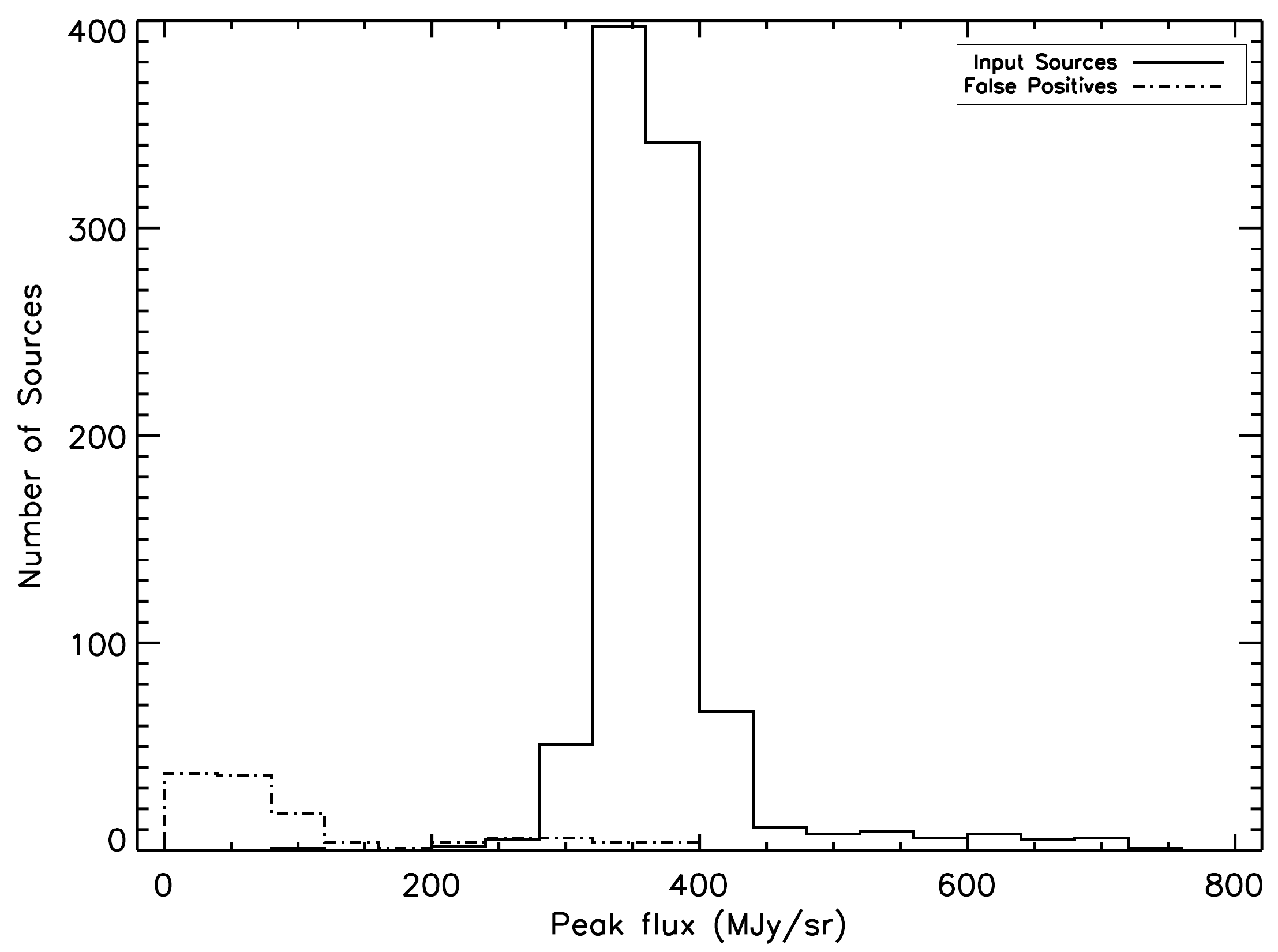}}
\resizebox{\hsize}{!}{\includegraphics{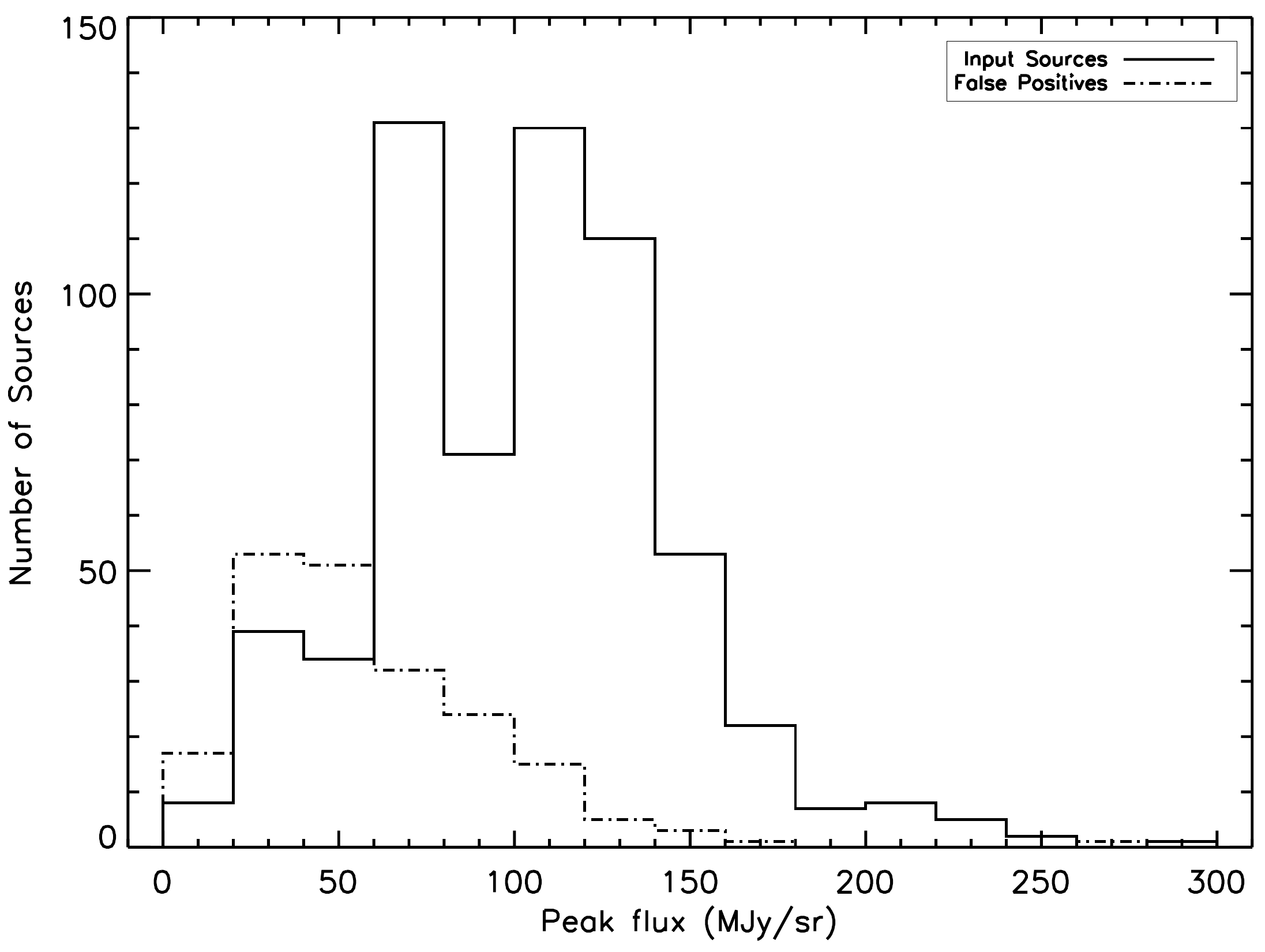}}
\caption{Histograms for the distributions of integrated fluxes for the \textit{false positive} sources which were recovered in the simulations for F$_{peak}$=0.3 Jy/pxl sources (top panel) and for the F$_{peak}$=0.1 Jy/pxl sources (bottom panel). In each figure the full line represent the distribution of input fluxes, while the dot-dashed line is the distribution of recovered fluxes for the false positives. The results reported are those for a curvature threshold of 0.5\sigdder.}
\label{false_pos}
\end{figure}

Lower curvature detection thresholds ensure higher source detection rates as shown in Fig. \ref{det_results}. This, however, causes an increasing fraction of spurious detections (false positives). This may not be a big problem, in principle, as the peak fluxes estimated for these false positives are lower compared to the fluxes for the positively matched sources (see Fig. \ref{false_pos}) and, as such, most of them can be easily spotted and removed. Fig. \ref{false_pos} shows that false positives are at the level of the noise of the simulated map (50 MJy/sr roughly corresponds  to 45mJy/pxl in the 6\asec\ pixel of the simulations). 

We caution, however, that this is the situation we find in these simulations, with this assumed structure of underlying diffuse background. Other choices of background spatial morphological properties might in principle lead to different flux levels for false positives; in other words we might start picking up statistical fluctuations of the ISM structure. This is more an astronomical problem than a pure detection problem. The prospective users of this algorithm who may wish to employ it in notably different background conditions with respect to the test conditions adopted in the present article are advised to carry out their own checks by cross-correlating detections at nearby wavelengths to confirm the real nature of a structure as opposed to a statistical background fluctuations, but especially performing custom artificial source experiments like the ones we have described.

\begin{figure}[h]
\resizebox{\hsize}{!}{\includegraphics{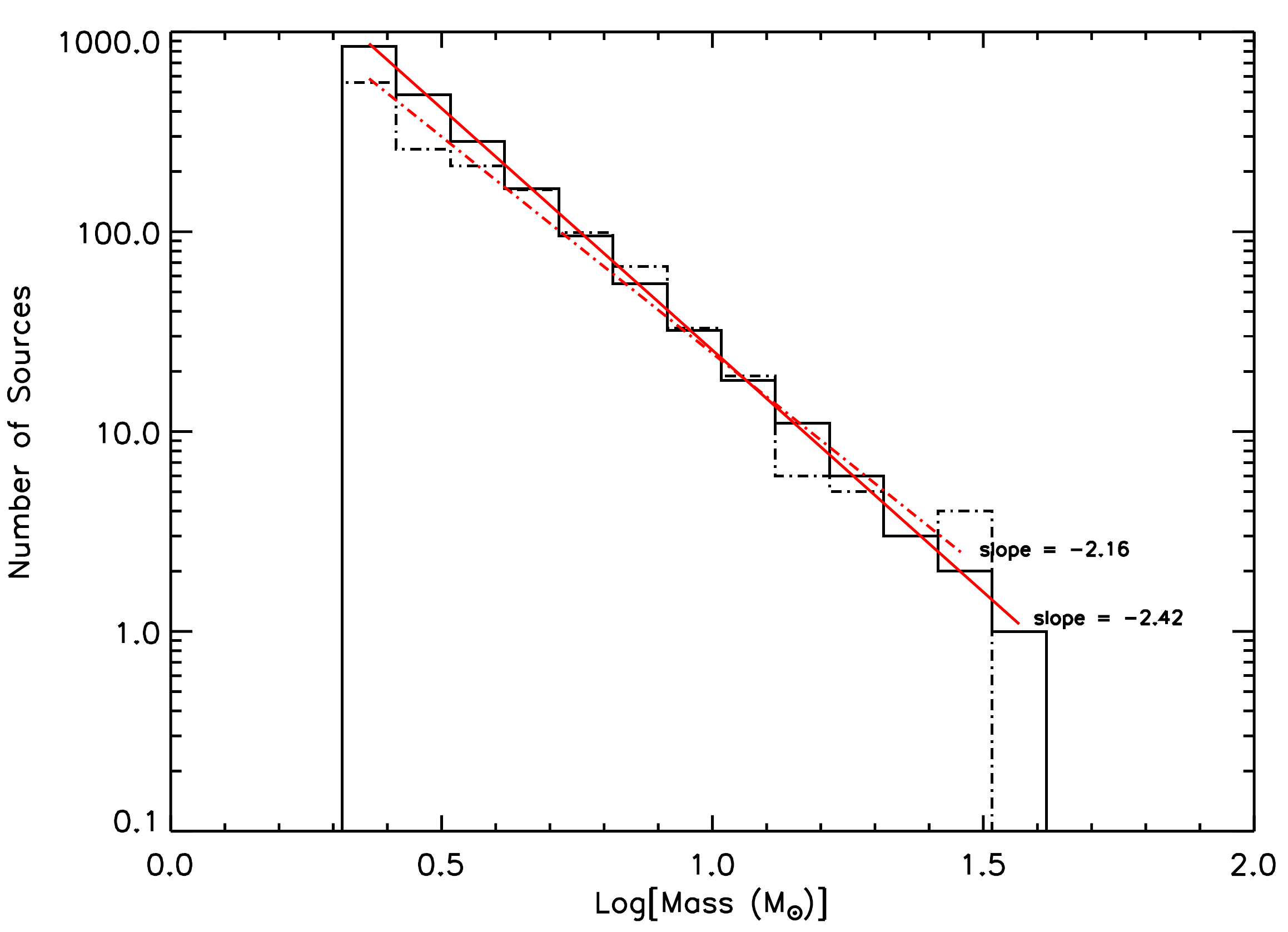}}
\caption{Mass functions for a simulation of 2000 sources with mass extracted from a Salpeter IMF (full black line). The full red line is a linear fit to the simulated mass function, where the slope is -2.42$\pm$0.02. The dash-dotted line is the mass distribution of the sources extracted form the simulated field. The red dash-dotted line is the linear fit to the distribution; the recovered mass function slope is 2.2$\pm$0.1, and it agrees with the input one within a 2$\sigma$ uncertainty. }
\label{imf_check}
\end{figure}

In each of the four test cases above, the peak flux of all sources was kept constant to get a reliable estimate of completeness levels for the extracted source catalogues. An additional test case was also generated to test the ability of the software to retrieve an astronomically significant result, like the mass function of cold cores. A 1\adeg\ x1\adeg field was generated with the same realization of diffuse emission with the white noise component as for the previous tests, but injecting this time 2000 sources whose mass was extracted from a \citet{salpeter55} IMF. The resulting input mass spectrum is indicated  in Fig. \ref{imf_check}  by the full line histogram; the slope as measured from a fit to the distribution results is -2.42$\pm$0.02. Masses were converted into fluxes assuming a dust temperature of 20K and dust opacity from \citet{preib93} with $beta=1.5$. These fluxes were assumed as integrated values for the sources. These were then distributed over the simulated field with exactly he same procedure as for the previous simulations, including the same random distribution of source sizes and ellipticities. The resulting peak fluxes have a lower boundary in the 200 mJy/beam regime, comparable to one of the four test cases described before.

The source extraction was carried out with a \textit{curvature} of 0.5\sigdder, and retrieved 1690 out of the 2000 injected sources. The integrated fluxes were converted back into mass using the same assumptions used to generate the input sources. The distribution of recovered masses is reported in Fig. \ref{imf_check} with the dash-dot histogram; a fit to this distribution (red dash-dot line) yields a slope of -2.2$\pm$0.1. The recovered slope can be reconciled with the input one within a 2$\sigma$ uncertainty.

\subsection{Fine-tuning of parameters and their impact on algorithm performance}

We conclude by analyzing the dependence of the CUTEX performances on a few parameters that were kept fixed in the above simulations.

\subsubsection{Image pixel scale}
\label{pixscale}

Code development and test was carried out using simulated images with pixel sizes equal to $1/3$ of the beam FWHM at the given simulation wavelength. This formally implies a higher frequency as opposed to the commonly quoted Nyquist sampling theorem that, on the other hand, was developed for 1D sine-wave signals. In 2D, the image quality improves significantly with smaller pixel sizes (in particular, the $1/3$-sampling of the beam FWHM would ensure that the Nyquist sampling is achieved also on diagonal pixel directions), provided that one has enough signal redundancy to avoid losses of final S/N due to poor coverage. Additional tests were carried out on simulated images with a coarser pixel size (half the beam FWHM) using lists of 1000 synthetic sources with peak fluxes of 100 and 300 mJy/beam, which were the two lowest flux regimes explored in the simulations. We find a significant decrease of the performances in terms of correct flux retrieval only for the faintest objects (that are at an equivalent 2.5$\sigma$ flux levels), while the performances with 300mJy/beam sources are unchanged. This coarser sampling causes the fainter point-like objects to appear like low level spikes, with one pixel only marginally emerging from the noise, and the algorithm clearly fails in these conditions.

\subsubsection{Relationship between curvature detection threshold and $\tilde n_{CMP}$}
\label{cmp_members}

To characterize the relationship between the curvature detection threshold and the number of CMPs in a CMP cluster we used a simple simulation consisting in 1 source the size of the PSF; pixel-to-pixel random noise at different levels was added to simulate different source peak S/N values of 5, 10 and 30. For each of these three cases we ran the CUTEX detection stage with 5 different curvature detection thresholds of $\zeta _{th}=$0.25, 0.5, 0.75, 1.0 and 2.0\sigdder . We find that in all cases but $\zeta _{th}$=0.25\sigdder, the value of $n_{CMP}$ for the cluster at the source location is always strictly greater than 2, while the CMP clusters originated by the noise elsewhere in the simulated image always have $n_{CMP}\leq 2$. Curvature detection threshold $\zeta _{th} \leq 0.25$\sigdder\  are found to generate several CMP clusters with 3 pixels, that would then result in false-positive source detections. We then regard $\tilde{n} _{CMP}=3$ as the minimum for a CMP cluster to qualify as a positive candidate source to be adopted, for curvature detection thresholds $\zeta_{th} \geq 0.5$\sigdder. We stress, however, that the parameter $\tilde{n} _{CMP}$ can be set to a non-default value in case the specific case at hand requires it (e.g., peculiar background conditions different from the ones we used in our simulations).

\subsubsection{Source grouping criteria}
\label{tune_thresh}

The reciprocal distance threshold that we adopted for simultaneous Gaussian fitting is twice the initial guess of the source FWHM, and it is a tunable parameter in CUTEX. The default adopted value is principle more than sufficient to account for reciprocal source contamination. Additional simulations were carried out with 1000 F$_{peak}$=0.3Jy synthetic sources distributed as in Fig. \ref{sim_fields}, varying the grouping threshold between 1.5 and 3 times the PSF. We report no substantial changes in the overall algorithm performances.

Threshold values below 1.5 times the PSF should not be used given that the whole point is to estimate mutual contamination; it is also not advisable because a relatively close companion and fainter companion will perturb the convergence of the Gaussian fit and lead to incorrect final source positioning. Grouping thresholds larger than our adopted default value may lead (especially in extreme crowding conditions) to large groups, where mutual contamination is no longer an issue and simultaneous fitting is not really justified.

\subsubsection{Constraints on source size fitting}
\label{size_assum}

\begin{figure}[t]
\resizebox{\hsize}{!}{\includegraphics{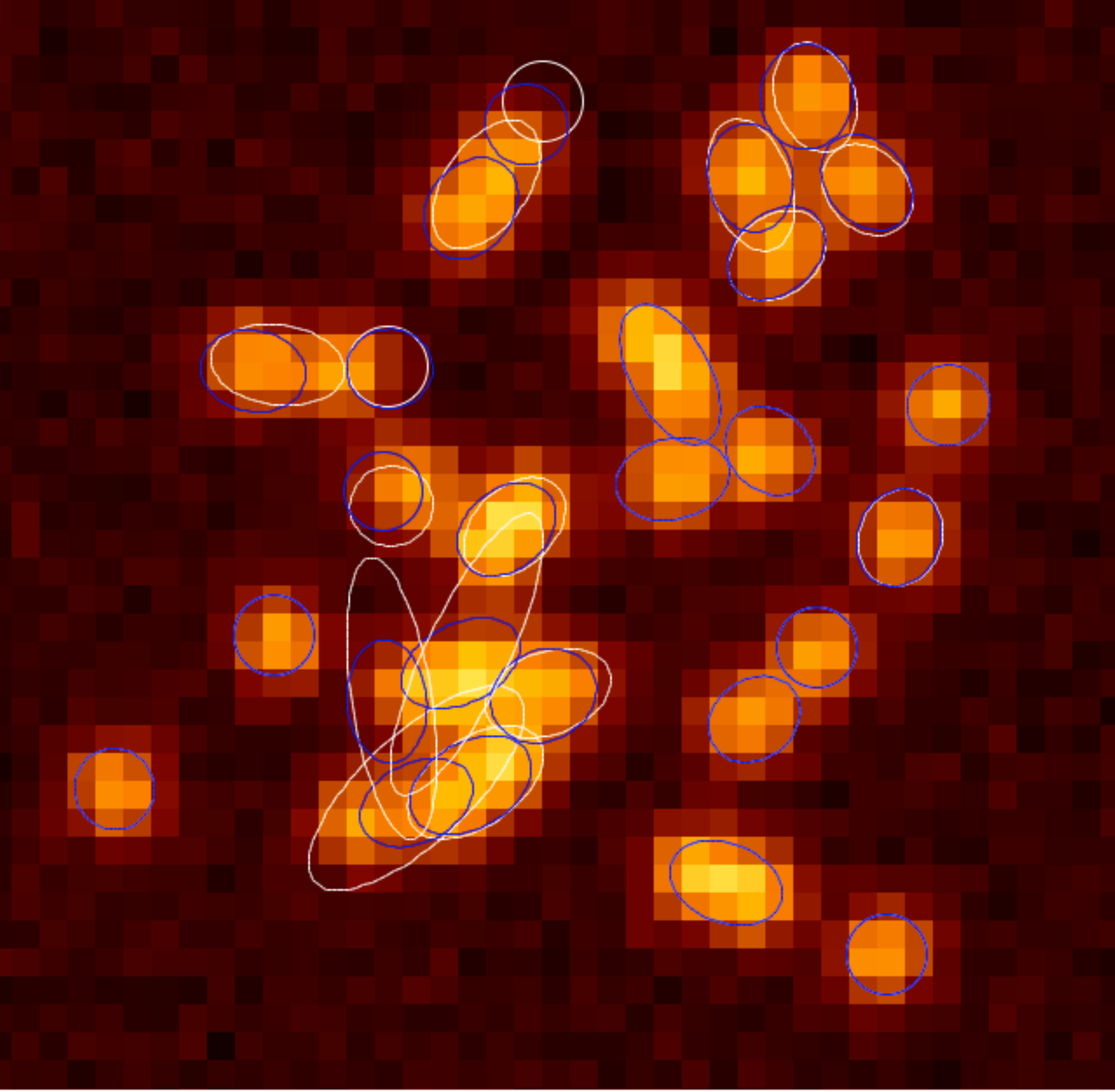}}
\caption{Simulation of 30 closely packed PSF-sized objects. The blue ellipses are the sources estimated with the $\pm$20\% size constraint on the Gaussian fitting. The white ellipses are the same sources estimated with the sizes as free parameters. When white ellipses cannot be seen is because the exactly overlap with the blue ones.}
\label{crowded_field}
\end{figure}

In our Gaussian fitting we allow the source sizes to vary by 20\% at most with respect to the initial values estimated from the detection stage. Additional tests were carried out using the same simulated fields as in Fig. \ref{sim_fields} and by letting the source sizes as free parameters in the fit; no notable degradation in performances has been observed. We also created a simulated field with 30 compact sources closely packed, and checked the impact of leaving the source sizes parameters as free in the Gaussian fit. As reported in Fig. \ref{crowded_field}, notable differences are found in only a few cases where the packing of sources is really critical. Most of the sources are not affected. This reassures us that imposing the size constraints does not introduce biases in the flux estimates, and is useful to help reasonable convergence in extremely crowded regions.

\section{Conclusions}

We have developed a simple code called CUTEX (CUrvature Thresholding EXtractor) implemented in IDL language to approach in a novel way the problem of source detection and photometry in relatively crowded fields with variable-size sources embedded in spatially variable background. These conditions are the most demanding for the performances of many of the most commonly used packages.

Our detection method is based on the analysis of the curvature properties of the astronomical image, rather than on properties of the signal intensity. Multidirectional double-differentiation yields a map of the curvature of the intensity distribution, greatly enhancing compact sources and erasing the contribution from diffuse emission without having to resort to matched filtering with fixed source "templates". Thresholding and subsequent segmentation of high-curvature areas of the image yields candidate source positions. By detecting changes of sign in the curvature, it is relatively easy to detect where each source merges into a morphologically differing structure, therefore greatly aiding the separation of, e.g., core-plateau structures. This information is used to estimate accurate source sizes. We also find that changes in the second derivatives of the signal are very effective in spotting blended sources. 

Subsequent photometry estimate is performed by fitting elliptical 2D Gaussian plus a planar plateau to all detected peaks.  Blended peaks are fit with more Gaussians simultaneously. A meaningful and reliable convergence is aided by an accurate estimate of the initial sources size which allow us to keep quasi-fixed parameters. When multiple peaks are fit the distance among the peaks is kept fixed so that the gaussians are fit "as a group" rather than independently.

CUTEX has been tested over a suite of simulated images with a realistic highly-variable diffuse emission as background, and populated with thousands of synthetic sources of variable size and intensity. We find source detection rates in excess of 90\%\ for a few $\sigma$ equivalent threshold, with a very encouraging flux recovery accuracy except for resolved sources with peak flux below 3$\sigma$.

We plan to apply CUTEX to the source extraction for the major \textit{Herschel} satellite Galactic imaging surveys, starting with Hi-GAL \citep{moli10a}. The code is not yet in a form compatible with public release, but we plan to make it freely available in a year timescale. A beta-version can be requested to the authors.

We conclude by stressing again that the prospective users of this algorithm are strongly encouraged to carry out tests with artificial source experiments like the ones we have described in the paper, to verify that the default algorithm parameters work fine for their needs or to determine a different set of parameters (e.g. $\tilde{n}_{CMP}$, source grouping threshold, etc.).

\begin{acknowledgement}
The initial development and testing of the source photometry part of the code was carried out during a visit of S.M. at the Spitzer Science Center, whose kind hospitality is here acknowledge. We particularly thank A. Noriega-Crespo and S. Carey for suggesting the use of the MPFIT package, which greatly helped our efforts, and for continuing discussions and advise about code performance and test results. We thank the referee, Dr. Jonathan Williams, for valuable suggestions and comments that led to an improved code and paper clarity. We also thank D. Elia for providing the code for fractal-based realization of 2D images.
\end{acknowledgement}

\bibliography{/Users/molinari/science/sergio_bib}
\bibliographystyle{aa}

\end{document}